\documentclass[pre,aps,twocolumn,superscriptaddress]{revtex4-2}

\usepackage[english]{babel} 
\usepackage{graphicx}
\usepackage{amssymb,amsfonts,amsmath}
\usepackage{color}
\usepackage[hidelinks]{hyperref}
\usepackage[normalem]{ulem}
\usepackage{natbib}

\usepackage{tikz}
\usetikzlibrary{shapes}
\usetikzlibrary{patterns}
\usetikzlibrary{angles, quotes}
\definecolor{bostonuniversityred}{rgb}{0.8, 0.0, 0.0}
\definecolor{chromeyellow}{rgb}{1.0, 0.65, 0.0}

\newcommand{\rv}{{\mathbf r}}
\renewcommand{\vec}{\mathbf}

\newcommand{\red}{\color{black}}

\usepackage{amssymb}

\usepackage[final]{showlabels} 

\begin{document}

\title{\red Combining integral equation closures with force density functional theory\\ for the study of inhomogeneous fluids
}

\author{S.~M. Tschopp}
\affiliation{Department of Physics, University of Fribourg, CH-1700 Fribourg, Switzerland}
\author{H. Vahid}
\affiliation{Leibniz-Institute for Polymer Research, Institute Theory of Polymers, D-01069 Dresden, Germany}
\author{A. Sharma}
\affiliation{University of Augsburg, Institute of Physics, D-86159 Augsburg, Germany}
\author{J.~M. Brader}
\affiliation{Department of Physics, University of Fribourg, CH-1700 Fribourg, Switzerland}
\date{\today}

\begin{abstract}
Classical density functional theory (DFT) is a powerful framework to study inhomogeneous fluids. Its standard form is based on the knowledge of a generating free energy functional. 
If this is known exactly, then the results obtained by using standard DFT or its alternative, recently developed 
version, force-DFT, are the same. 
If the free energy functional is known only approximately then these two routes produce different outcomes. 
However, as we show in this work, force-DFT has the advantage that it is also implementable without knowledge of the 
free energy functional, 
by using instead liquid-state integral equation closures. 
This broadens the range of systems that can be explored, since free energy functionals are generally difficult 
to approximate.
In this paper we {\red investigate} the utility of using inhomogeneous integral equation closures within force-DFT 
thus demonstrating the versatility and accuracy of this approach.
\end{abstract}

\maketitle

\section{Introduction}

Classical density functional theory (DFT) is a powerful and well-established framework for the study of inhomogeneous 
fluids in equilibrium \cite{evans79,evans92}.
Traditional applications of DFT require knowledge of the excess Helmholtz free energy functional, which contains 
all nontrivial information about the interparticle interactions.  
Functional differentiation then generates the one-body direct correlation function required to solve the Euler-Lagrange (EL) equation for the one-body density. 
Unfortunately, practical implementation of this standard 
DFT scheme is often hindered by the limited availability of 
reliable and accurate approximations to the excess Helmholtz free energy functional.
While quantitatively accurate approximations do exist for  
both hard- and penetrable-particle models (using fundamental 
measures theory and density expansion functionals \cite{evans92,RothReview}, respectively) the 
situation is less comfortable when considering either particles with 
{\red a strongly repulsive soft core or those with an attractive component to the 
interaction potential (although  progress has recently been made in the latter case \cite{tschopp1,tschopp2}, improving upon established approximations 
\cite{articleDFTbetter,RosenfeldBridge}). 
}
  
The recently proposed force-DFT approach \cite{tschopp3} 
provides an alternative way to obtain one-body density profiles from a given excess Helmholtz free energy functional, by explicitly treating the inhomogeneous two-body correlation functions.
Working on the two-body level provides higher resolution of the fluid's internal microstructure. This feature enables direct calculation of the average interparticle forces acting within the system via spatial integration and thus provides access to physical quantities such as the surface tension \cite{widom}. 
By identifying the inhomogeneous two-body correlations as functionals of the one-body density \cite{tschopp1,tschopp2}, the force-DFT approach retains the spirit of the traditional DFT framework.
If the Helmholtz free energy functional is known exactly, these two approaches become strictly equivalent.

The inhomogeneous correlation functions used in force-DFT are obtained via solution of the inhomogeneous Ornstein-Zernike (OZ) equation.
This feature enables versatility in the choice of an implementation route. The one presented in Reference \cite{tschopp3} did indeed exploit the second functional derivative of the excess Helmholtz free energy to obtain the two-body direct correlation function and thus permit the solution of the inhomogeneous OZ equation. However, from the field of liquid-state integral equation theory, it is known that the OZ equation can be solved by supplementing it with {\red an approximate} closure relation \cite{hansen06,caccamo,Bomont,Santos}.
This second approach does not involve the knowledge of an excess free energy functional, but still retains the feature that the two-body correlations are functionals of the one-body density (since the latter remains {\red the sole} input to the 
{\red inhomogeneous} OZ equation). 
{\red Information regarding the interparticle interaction potential is contained within the closure relation.}

Long experience with closures of the bulk OZ equation has lead to a selection of reliable approximations for the pair correlation functions which, if chosen wisely, can yield accurate results for a broad range of relevant interparticle interaction potentials. 
{\red For example, hard or other strongly repulsive potentials 
can be accurately treated using closures due to 
Verlet \cite{verlet_closure}, Martynov and Sarkisov 
\cite{martynov_closure}, 
Rosenfeld and Ashcroft \cite{MHNC} or 
Rogers and Young \cite{rogersyoung}, whereas attractive interactions are well accounted for 
by the Duh-Haymet closure 
\cite{DuhHaymet1,DuhHaymet2} or, in certain cases, 
the powerful self-consistent Ornstein Zernike 
approximation (SCOZA) of Stell and coworkers 
\cite{scozaOther,scozaWilding}. 
Since these references represent only the tip of the iceberg regarding the world of integral equation closures, we direct the interested reader 
to the classic article of Barker and Henderson \cite{BarkerHendersonLiquid} and the more recent 
work of Pihlajamaa and Janssen \cite{Janssen} 
for an overview and assessment of the various approximations available.}
In any case, it is fair to say that the number of systems which can be accurately treated in bulk using liquid-state integral equation theory well exceeds the number of systems for which we have a good approximation to the excess Helmholtz free energy functional.
It is therefore of interest to investigate the possibility of generalising those closure relations to treat inhomogeneous systems.

In the present paper we choose to focus on a system of soft repulsive particles in two-dimensions, for which the excess Helmholtz free energy functional has at present no truly accurate approximation, but for which there exist a pool of accurate closure relations. 
We will thereby show that the new route to calculating density profiles using force-DFT presents a powerful alternative to the standard DFT scheme.

\section{Theory}

\subsection{Density functional theory in a nutshell}

\subsubsection{Standard DFT as a reminder}

The classical DFT is an exact formalism for the study of 
many-body systems in external fields \cite{evans79,evans92}. In its standard form, 
the central object of interest is the grand potential functional
\begin{equation}\label{grand potential}
\Omega[\,\rho\,] = F^{\text{id}}[\,\rho\,] + F^{\text{exc}}[\,\rho\,] 
- \int \!d\rv \, \big( \mu - V_{\text{ext}}(\rv) \big)\rho(\rv), 
\end{equation}
where $\mu$ is the chemical potential, $V_{\text{ext}}$ is the external potential and 
$\rho$ is the ensemble averaged one-body density.
The Helmholtz free energy of the ideal gas is exactly given by	
\begin{equation}\label{Helmholtz free energy}
F^{\text{id}}[\,\rho\,]=k_BT\int\!d\rv\, \rho(\rv) \big( \ln(\rho(\rv))-1 \big), 
\end{equation}
where $k_B$ is the Boltzmann constant, $T$ is the temperature and, without loss of generality, 
we have set the thermal wavelength equal to unity. 
The excess Helmholtz free energy, $F^{\text{exc}}$, includes all information 
regarding the interparticle interactions and usually has to be approximated.
The grand potential satisfies the following variational condition \cite{evans79}
\begin{align}
\label{variational principle}
\frac{\delta  \Omega[\,\rho\,]}{\delta \rho(\rv)}{\bigg|}_{\rho=\rho_{\text{eq}}}=0, 
\end{align}
which yields the EL equation,
\begin{align}\label{EL equation}
\rho(\rv)=e^{ -\beta \, \big(V_{\text{ext}}(\rv) \,-\, \mu \,-\, k_BT \, c^{(1)}(\rv) \big)},
\end{align}
where $\beta\!=\!1/(k_BT)$ and the one-body direct correlation function, $c^{(1)}$, is generated from the excess 
Helmholtz free energy by a functional derivative,
\begin{align}\label{c1 definition}
c^{(1)}(\rv)=-\frac{\delta \beta F^{\text{exc}}[\,\rho\,]}{\delta\rho(\rv)}{\bigg|}_{\rho=\rho_{\text{eq}}}.
\end{align}
Substitution of the solution of \eqref{EL equation} into \eqref{grand potential} yields the equilibrium 
grand potential, and thus provides access to all thermodynamic properties of 
the system.
It is clear that this framework is practically useful only when the excess Helmholtz free energy functional, $F^{\text{exc}}$, is known.

\subsubsection{Force-DFT as chosen framework}

Force-DFT \cite{tschopp3} is an alternative route to obtaining inhomogeneous density profiles 
for systems interacting via a pairwise additive interparticle potential, $\phi$.
In this framework, the main equation is no longer the EL equation \eqref{EL equation}, but the Yvon-Born-Green (YBG) equation, 
\begin{align} \label{YBG equation}
0 = \nabla_{\vec{r}_1}  \rho(\vec{r}_1) &+  \rho(\vec{r}_1) \nabla_{\vec{r}_1} \beta V_{\text{ext}}(\vec{r}_1) \\
&+ \!\int\! d \vec{r}_2 \,  \rho^{(2)}(\vec{r}_1,\vec{r}_2) \nabla_{\vec{r}_1} \beta \phi(r_{12}), \notag
\end{align}
where $r_{12}\!\equiv\!|\vec{r}_1-\vec{r}_2|$. 
The YBG equation expresses the balance between the Brownian, the external and the internal forces required at equilibrium \cite{hansen06}. It is important to highlight that the two-body density, $\rho^{(2)}$, appearing in equation 
\eqref{YBG equation} is a functional of the one-body density \cite{tschopp1, tschopp2} (as was the one-body direct correlation function, $c^{(1)}$, in the standard variational scheme). 
While previously expression \eqref{c1 definition} was sufficient to obtain $c^{(1)}$ and close the EL equation, obtaining $\rho^{(2)}$ as a functional of $\rho$ is a more demanding 
task, since we are now working on the two-body level. At this stage, it may be worth mentioning that the force-DFT framework holds regardless of the specific scheme chosen to obtain the equilibrium two-body density, as long as it remains a functional of the inhomogeneous one-body density. 
{\red The force-DFT cannot be applied to systems interacting via triplet- or higher-body interactions as 
a consequence of working on the level of the two-body correlation functions.}

The most effective route to generating two-body correlation functions (which are actually all functionals of the one-body density) consists of solving the inhomogeneous OZ equation,
\begin{equation} \label{OZ equation}
h(\rv_1,\rv_2) = c(\rv_1,\rv_2) 
+ \!\int\!\! d\rv_3\, h(\rv_1,\rv_3)\,  \rho(\rv_3) \, c(\rv_3,\rv_2),
\end{equation}
where $c$ is the two-body direct correlation function, also given by
\begin{equation} \label{c2 definition}
c(\rv_1,\rv_2)=-{\frac{\delta^2 \beta F^{\text{exc}}[\rho]}{\delta\rho(\rv_1) \delta\rho(\rv_2)}}\bigg\rvert_{\rho=\rho_{\text{eq}}}, 
\end{equation}
when the excess Helmholtz free energy functional, $F^{\text{exc}}$, is known, and $h$ is the two-body total correlation function related to $\rho^{(2)}$ via
\begin{equation} \label{rho2 definition}
h(\rv_1, \rv_2) = \frac{\rho^{(2)}(\rv_1, \rv_2)}{ \rho(\rv_1)  \rho(\rv_2)} -1.
\end{equation}

If one has access to $F^{\text{exc}}$, equations \eqref{YBG equation},\eqref{OZ equation}, \eqref{c2 definition} and \eqref{rho2 definition} form a closed set and yield the force-DFT as introduced in Reference \cite{tschopp3}.
In the special case that the excess Helmholtz free energy functional is known exactly, the one-body density profile generated by this scheme is completely equivalent to the one obtained by standard DFT, via equations \eqref{EL equation} and \eqref{c1 definition}. If $F^{\text{exc}}$ is only known approximately then this is no longer the case. The standard DFT one-body density profiles are consistent with the thermodynamical quantities associated with the compressibility route, while the force-DFT profiles are consistent with the virial route \cite{tschopp3}.

However, there is no need to know the excess Helmholtz free energy functional in order to solve the inhomogeneous OZ equation \eqref{OZ equation}. One can replace equation \eqref{c2 definition} by a closure relation, in the spirit of liquid-state integral equation theory. It is this implementation scheme that we will use in the present work. 

In principle, knowledge of both the inhomogeneous one- and two-body densities gives access to all thermodynamic quantities via integration. In that respect force-DFT is not less versatile than standard DFT. If one is interested in the internal micro-structure of the inhomogeneous fluid, then having explicit access to $\rho^{(2)}$ is clearly advantageous.

\subsection{Specification of the interparticle interaction potential}

In this work, we choose to focus on (two-dimensional) hard-core Yukawa particles by setting
\begin{align}\label{HCY potential}
\phi(r_{12}) = 
\begin{cases} 
      \hspace*{0.9cm}\infty, & r_{12} < d, \vspace*{0.2cm}\\
      \kappa\,\frac{e^{-\alpha(r_{12}-d)}}{r_{12}}, & r_{12}\ge d,
\end{cases}
\end{align}
where the particle (hard-core) diameter $d\!=\!1$.
We fix the amplitude to the (positive) value $\kappa\!=\!10$, but leave the inverse decay length $\alpha$ as a parameter.


{\red This type of particles are interesting because they have a hard core preventing overlaps, but are softened 
by a tunable repulsive Yukawa tail.} 
The following exact relation
\begin{equation}\label{hard-core condition}
g(\vec{r}_1,\vec{r}_2) \equiv h(\vec{r}_1,\vec{r}_2) +1 = 0, \qquad r_{12}<d,
\end{equation}
thus applies.
There is no reliable free energy functional available for such systems. Although perturbation theory can provide accurate results when applied 
to treat attractive interactions \cite{tschopp1,tschopp2,articleDFTbetter}, it fares less well for a repulsive soft tail \cite{BarkerHendersonLiquid}. Standard DFT may thus not be the best way to proceed.
However force-DFT is suitable for treating these softened particles since there exist a collection of closures known to perform well on such systems \cite{BarkerHendersonLiquid,Janssen}.

\subsection{Specification of the closure relations}

Using the framework of integral equations and generalising it to inhomogeneous systems, the two-body correlation function is formally given by
\begin{equation}
g(\vec{r}_1,\vec{r}_2) = e^{-\beta \phi(r_{12})} e^{\gamma(\vec{r}_1,\vec{r}_2)} e^{b(\vec{r}_1, \vec{r}_2)} ,
\end{equation}
where we defined the continuous function $\gamma\!\equiv\!h-c$ and where $b$ is the bridge function. This latter quantity is generally unknown but can be approximated using a closure relation.

{\red

\subsubsection{The Hypernetted chain approximation}

In some sense
the simplest closure is the Hypernetted chain (HNC) approximation, that consists in setting the bridge function to zero \cite{Janssen}, i.e.
\begin{equation}\label{HNC bridge}
b_{\text{HNC}}(\vec{r}_1, \vec{r}_2) = 0.
\end{equation}
This is equivalent to imposing
\begin{equation}\label{HNC closure}
\!\! c(\vec{r}_1, \vec{r}_2) \!\overset{\text{HNC}}{=}\! e^{\!-\beta \phi(r_{12}) + \gamma(\vec{r}_1, \vec{r}_2)} \!-\! \gamma(\vec{r}_1, \vec{r}_2) \!-\! 1.
\end{equation}
This approximation was originally derived in References 
\cite{HNCoriginal1,HNCoriginal2} and is known to perform well for soft penetrable 
particles, but is less reliable for strongly repulsive systems.
}

\subsubsection{The Percus-Yevick approximation}

The generalised Percus-Yevick (PY) closure is given by
\begin{equation}
b_{\text{PY}}(\vec{r}_1, \vec{r}_2) = \ln(\gamma(\vec{r}_1, \vec{r}_2)+1) - \gamma(\vec{r}_1, \vec{r}_2).
\end{equation}
Since $g\!\equiv\!h+1$, this is equivalent to imposing
\begin{equation}\label{PY closure}
c(\vec{r}_1, \vec{r}_2) \overset{\text{PY}}{=} \left( e^{-\beta \phi(r_{12})} -1 \right) (\gamma(\vec{r}_1, \vec{r}_2)+1).
\end{equation}

For a system of purely hard particles (hard-disks in two dimensions) this reduces to the well-known relation
\begin{equation}\label{HD PY closure}
\begin{cases}
h(\vec{r}_1, \vec{r}_2) = -1, \qquad r_{12}<d ,\\
c(\vec{r}_1, \vec{r}_2) = 0, \qquad \quad r_{12}>d.
\end{cases}
\end{equation}
The first line recovers the exact condition \eqref{hard-core condition}, while the second line \textit{is} the hard-disk PY approximation.
{\red This closure was originally proposed in Reference 
\cite{PercusOriginal} and provides good results for strongly repulsive interparticle interactions.
}

\subsubsection{The Verlet approximation}

Another closure relation for repulsive particles is the Verlet (V) closure, where the bridge function is defined by \cite{Janssen,verlet_closure}
\begin{equation}\label{V bridge}
b_{\text{V}}(\vec{r}_1, \vec{r}_2) = - \frac{\frac{1}{2} \gamma^2(\vec{r}_1, \vec{r}_2)}{1+\frac{4}{5} \gamma(\vec{r}_1, \vec{r}_2)}.
\end{equation}
This is equivalent to imposing
\begin{equation}\label{V closure}
\!\! c(\vec{r}_1, \vec{r}_2) \!\overset{\text{V}}{=}\! e^{\!\!-\beta \phi(r_{12}) + \gamma(\vec{r}_1, \vec{r}_2)- \frac{\frac{1}{2} \gamma^2(\vec{r}_1, \vec{r}_2)}{1+\frac{4}{5} \gamma(\vec{r}_1, \vec{r}_2)}} \!-\! \gamma(\vec{r}_1, \vec{r}_2) \!\!-\!\! 1.
\end{equation}
The form of the bridge function 
\eqref{V bridge} was chosen by Verlet to exactly reproduce the PY virial coefficients up to fourth-order in the bulk 
density. The fifth- and higher-order virial coefficients 
predicted by equation \eqref{V bridge} are then found to improve considerably on the PY values.
{\red This closure relation was created in order to improve upon the PY approximation for three-dimensional hard spheres.
Note that its performance for systems with attractive interactions has not yet been fully investigated.
}

\subsubsection{The Martynov-Sarkisov approximation}

Last but not least, the Martynov-Sarkisov (MS) closure is given by \cite{Janssen,martynov_closure}
\begin{equation}\label{MS bridge}
b_{\text{MS}}(\vec{r}_1, \vec{r}_2) = \sqrt{1+2 \gamma(\vec{r}_1, \vec{r}_2)} - \gamma(\vec{r}_1, \vec{r}_2) - 1,
\end{equation}
which is equivalent to imposing
\begin{equation}\label{MS closure}
c(\vec{r}_1, \vec{r}_2) \overset{\text{MS}}{=} e^{-\beta \phi(r_{12}) + \sqrt{1+2 \gamma(\vec{r}_1, \vec{r}_2)} - 1 } - \gamma(\vec{r}_1, \vec{r}_2) - 1.
\end{equation}
The approximation \eqref{MS bridge} was derived by expanding the 
bridge function in powers of $\omega\!\equiv\!\gamma\!+\!b$, which Martynov and co-workers (but apparently nobody else) named the `thermal potential'. Truncation of this expansion at leading order in $\omega$ and 
solution of the resulting quadratic equation for $b$ then 
leads to equation \eqref{MS bridge}.
{\red A major drawback of this approximation is that the presence of a square root in equation \eqref{MS closure} can lead to a breakdown of the theory in the event that the quantity $1\!+\!2 \gamma$ becomes negative.
}

\subsection{Specification of the external fields and geometry}

\begin{figure}[!t]
\hspace*{-0.5cm}
\begin{center}
\begin{tikzpicture}
\coordinate (origine) at (0,0);
\coordinate (particle 1) at (0.85,0);
\coordinate (particle 2) at (1.4,1.45);
\coordinate (x axes upper end) at (0,2.75);
\coordinate (x axes lower end) at (0,-2.75);
\coordinate (z axes left end) at (-3.75,0);
\coordinate (z axes right end) at (3.75,0);
\coordinate (left wall lower end) at (2.25,-2.25);
\coordinate (left wall upper end) at (2.25,2.25);
\coordinate (left wall size) at (1,4.5);
\coordinate (right wall lower end) at (-2.25,-2.25);
\coordinate (right wall upper end) at (-2.25,2.25);
\coordinate (right wall size) at (-1,4.5);
\draw[-, >=latex, line width=0.8] (left wall lower end) -- (left wall upper end);
\draw[-, >=latex, line width=0.8] (right wall lower end) -- (right wall upper end);
\fill[pattern=north west lines] (left wall lower end) rectangle ++(left wall size);
\fill[pattern=north west lines] (right wall lower end) rectangle ++(right wall size);
\draw[->, >=latex, blue, line width=0.9] (x axes lower end) -- (x axes upper end);
\draw[->, >=latex, blue, line width=0.9] (z axes left end) -- (z axes right end);
\draw[blue] (x axes upper end) node[above] {\normalsize $x$};
\draw[blue] (z axes right end) node[right] {\normalsize $z$};
\filldraw (particle 1) circle (2.25pt);
\filldraw (particle 2) circle (2.25pt);
\draw (particle 1) node[above] {\small particle $1$};
\draw (particle 2) node[above] {\small particle $2$};
\draw[bostonuniversityred] (particle 1) node[below] {\normalsize ({\color{blue}$x_1 \!=\! 0$},$z_1$)};
\draw[bostonuniversityred] (particle 2) node[below] {\normalsize ($x_2$,$z_2$) };
\end{tikzpicture}
\end{center}
\caption{\textbf{Sketch of the used geometry and the relevant coordinates.}
We study a two-dimensional slit, using cartesian coordinates, where we replace the $y$-coordinate by $z$ to respect standard convention. 
}
\label{fig sketch geometry}
\end{figure}

So far, all above mentioned equations hold for any geometry and dimensionality. However, since we wish to focus on \textit{inhomogeneous} density profiles, we will impose an external field (and a dimensionality).

We choose to work in two dimensions, mainly since it is less well-studied than the three-dimensional case. This is largely due to the absence of accurate approximations to the excess Helmholtz free energy functional, $F^{\text{exc}}$, which has hindered standard DFT studies but poses no difficulties for the force-DFT scheme. Also an open two-dimensional system has the appealing feature that it still conserves the ensemble independence, which would become an issue in one dimension.

We focus on a slit, i.e. particles trapped between two (soft repulsive) walls. The external field is set as follows:
\begin{align}\label{external potential}
\! V_{\text{ext}}(z) = 
\begin{cases}
      \tilde{\kappa} \left( e^{\tilde{\alpha} (z - z_{w})} + e^{-\tilde{\alpha} (z + z_{w})} \right), & \!\! |z| < z_{w}, \vspace*{0.2cm}\\
      \hspace*{0.9cm}\infty, & \!\text{otherwise},
\end{cases}
\end{align}
where we defined $z_{w} \equiv \frac{L}{2}\!-\!\frac{d}{2}$, with the slit width $L\!=\!8$ and the particle diameter $d\!=\!1$. We fix the amplitude $\tilde{\kappa}\!=\!10$ as well as the inverse decay length $\tilde{\alpha}\!=\!3$.
Despite having set the problem in two dimensions we are using the coordinate $z$. We are actually taking the natural choice of cartesian coordinates $x$ and $y$, but we use the symbols $x$ and $z$ instead to respect the DFT community convention that usually employs the coordinate $z$ when working in planar geometry.

From the fundamental theorem of DFT \cite{evans79} we know that the external potential uniquely defines the so-induced one-body density profile and that they share the same geometrical symmetries. In our case, this means that the one-body density reduces to $\rho(z)$.
Since the one-body density does not vary in the $x$-direction, the two-body density is fully defined by $z_1$, $z_2$ and $x_{12}$, the distance between $x_1$ and $x_2$. Without lose of generality, we can choose to fix the coordinate axes such that $x_1\!=\!0$, as illustrated in figure \ref{fig sketch geometry}. This then yields $x_{12}\!=\!x_2$ and reduces the two-body density to $\rho^{(2)}(z_1,x_2,z_2)$.
We note that the distance between the center of particles one and two is now given by $r_{12}\!=\!\sqrt{x_2^2 + (z_1-z_2)^2}$. This is important when evaluating the pair potential \eqref{HCY potential}.

\subsection{Numerical implementation}

\subsubsection{Numerical implementation of the closure relation}

In our numerics, we chose to work with the two-body direct correlation function, $c$, and the continuous function $\gamma$ insead of the discontinuous two-body total correlation, $h$. This choice will become clearer in the next subsection as we will expose the need for repeated back and forth Fourier transformation and the related problem of using discontinuous functions.

The closure relations \eqref{PY closure}, \eqref{V closure} and \eqref{MS closure} hold for the whole domain. However, since they naturally reduce to the exact hard-core condition \eqref{hard-core condition} in the overlap-domain, we decided to explicitly treat these two cases separately as
\begin{equation*}\label{closure implementation}
c(z_1,x_2,z_2) \!=\!
\begin{cases}
-1 - \gamma(z_1,x_2,z_2), \qquad \qquad \qquad \, r_{12}<d ,\\
\text{\small closure relation \eqref{PY closure}, \eqref{V closure} or \eqref{MS closure}}, \; r_{12}>d,
\end{cases}
\end{equation*}
where we recall that $r_{12}\!=\!\sqrt{x_2^2 + (z_1-z_2)^2}$.

\subsubsection{Solving the inhomogenous OZ equation numerically}

The difficulty in solving the inhomogenous OZ equation \eqref{OZ equation} lies in its integral term.
In the case of a homogeneous system, the density $\rho(\vec{r}_3)$ reduces to a constant $\rho_b$ and can therefore be taken outside the integral. The integral term thus becomes a convolution. In this case, transforming to the Fourier-space helps to disentangle the spatial dependences and to solve the equation.
In the case of an inhomogeneous density it is more difficult to overcome the issue of the integral term. However, in specific geometries, namely the three-dimensional planar and spherical ones, it has been shown that it is possible to solve the OZ equation by transforming to the Hankel- or the Legendre-space, respectively \cite{tschopp1,tschopp2}.

In our present case of interest, the two-dimensional planar geometry, since the density only varies with respect to the coordinate $z$, we can use a one-dimensional Fourier transform in the $x$-dimension. As shown in Appendix \ref{appendix section FT OZ}, this yields the following equation:
\begin{align} \label{FT OZ equation planar eq1}
\tilde{h}(z_1,k,z_2) &= \tilde{c}(z_1,k,z_2) \\
&\quad+ \!\int_{-\infty}^{+\infty}\!\! dz_3\, \tilde{h}(z_1,k,z_3)\,  \rho(z_3) \, \tilde{c}(z_3,k,z_2), \notag
\end{align}
where $\tilde{h}$ and $\tilde{c}$ are the Fourier transforms of $h$ and $c$ with respect to the $x_2$-coordinate. Its equivalent form that uses $\tilde{\gamma}$ instead of $\tilde{h}$, namely
\begin{align} \label{FT OZ equation planar eq2}
\tilde{\gamma}(z_1,k,z_2) &= \!\int_{-\infty}^{+\infty}\!\! dz_3\, \tilde{c}(z_1,k,z_3)\,  \rho(z_3) \, \tilde{c}(z_3,k,z_2) \\
&\quad+\!\int_{-\infty}^{+\infty}\!\! dz_3\, \tilde{\gamma}(z_1,k,z_3)\,  \rho(z_3) \, \tilde{c}(z_3,k,z_2), \notag
\end{align}
is the one employed in this work. 
For numerical evaluation of the Fourier transforms, we followed the method of Lado \cite{Lado}.

Since the closure relation is set in real-space, the computational scheme needs an iterative back and forth Fourier transformation of the inhomogeneous two-body correlation functions, $\gamma$ and $c$, until convergence is reached.
The issue with this scheme is that discontinuous functions, such as $c$, are not suited to direct Fourier transformation due to the occurrence 
of Gibbs phenomena. This difficulty was already encountered in the three-dimensional case and is treated in Appendix C of Reference \cite{tschopp1} in the case of hard-spheres.

For our present problem, we use a similar scheme to overcome this issue.
First, as already stated, we are using the smooth function $\gamma$ instead of the discontinuous function $h$.
The discontinuous function $c$ remains to be dealt with.
The trick is to separate $c$ into a smooth part, that presents no issue to be Fourier transformed numerically, and a `step-slope' function that can be transformed analytically. When going back to real-space, the closure relation will help to reconstruct $c$ and $\gamma$. This technique rests on the fact that Fourier transformation is a linear operation.

Practically, this scheme is implemented as follows:
First we define $x_c\!\equiv\! \sqrt{d^2 + (z_1-z_2)^2}$, the critical $x_2$-value at which the discontinuity occurs.
Following the re-written closure relation from the last subsection, we know that inside the core the two-body direct correlation function is defined as
$c^{\text{in}}\!=\!-1-\gamma$, where $\gamma$ is a smooth function. There is thus no issue to employ interpolation tools on $\gamma$ and find the value and slope of $c$ at the inner discontinuity (i.e. the inner `jump' and `slope'), namely
\begin{align}
&c_{\text{jump}}^{\text{in}}(z_1,z_2) = -1 - \gamma(z_1,x_c,z_2), \\
&c_{\text{sl}}^{\text{in}}(z_1,z_2) = -\frac{\partial \gamma(z_1,x_2,z_2)}{\partial x_2}{\bigg|}_{x_2=x_c}.
\end{align}

In contrast to the aforementioned hard-spheres problem, our soft repulsive particles have a non-vanishing $c$-value outside the core. The present construction is thus a little more complex than in Reference \cite{tschopp1} where $c$ was set to zero outside the core and therefore there was no need to define the value and slope at the outer discontinuity.
In our soft case, $c$ is defined by the closure relation \eqref{PY closure}, \eqref{V closure} or \eqref{MS closure} outside the core. Since the whole aim of solving the inhomogeneous OZ equation is to determine this part of the two-body direct correlation function, we cannot use it to get the value and slope of $c$ at the outer discontinuity. However, luckily, there is a known first approximation to $c$ ouside the core, namely $c^{\text{out}} \!\approx\! -\phi$. This is part of the mean spherical approximation (MSA), a closure that is known to work well for attractive particles, but that simply \textit{is} the linearized PY closure \eqref{PY closure}. Since both the V and MS closures reduce to the PY closure for weak interparticle interactions, using $c^{\text{out}} \!\approx\! -\phi$ in our numerical treatment of the `jump' does not conflict with any of our chosen closures. This approximation, even if not that accurate, is good enough to smooth our numerical scheme. We thus use it to define the value and slope of $c$ at the outer discontinuity, as follows
\begin{align}
&c_{\text{jump}}^{\text{out}}(z_1,z_2) \approx -\phi(d) = -\kappa, \\
&c_{\text{sl}}^{\text{in}}(z_1,z_2) \approx \frac{-\phi\left(\sqrt{(x_c+\varepsilon_x)^2 + (z_1-z_2)^2}\right)+\kappa}{\varepsilon_x},
\end{align}
where, in the second line, $\varepsilon_x$ is the grid-spacing in the $x_2$-direction and we have used finite differences to obtain the total derivative of $\phi$ with respect to $r_{12}$ and evaluated at $d$.

Finally we are able to assign the `step' and (total) slope,
\begin{align}
&c_{\text{st}}(z_1,z_2) = c_{\text{jump}}^{\text{in}}(z_1,z_2) - c_{\text{jump}}^{\text{out}}(z_1,z_2),\\
&c_{\text{sl}}(z_1,z_2) = c_{\text{sl}}^{\text{in}}(z_1,z_2) - c_{\text{sl}}^{\text{out}}(z_1,z_2),
\end{align} 
which are the needed quantities to define the linear step-function,
\begin{equation}
f(z_1,x_2,z_2) = c_{\text{st}}(z_1,z_2)+c_{\text{sl}}(z_1,z_2) (x_2-x_c),
\end{equation}
for $x_2<x_c$ and zero otherwise,
that allows for the removal of the undesired discontinuity in $c$.

The procedure is quite straightforward and follows the same steps as in the end of Appendix C of Reference \cite{tschopp1}.
The linear step-function $f$ can be analytically Fourier transformed back and forth.
Defining the smoothed out correlation function
\begin{equation}
c_{\text{smooth}}(z_1,x_2,z_2) \equiv c(z_1,x_2,z_2)-f(z_1,x_2,z_2),
\end{equation}
allows for numerical transformations without too much extra noise.
Since, as already pointed out, the process of Fourier-transforming is additive, we can simply split $c$ into $c_{\text{smooth}}$ and $f$, separately transform them and then add them back together to get the Fourier transformed two-body correlation function, $\tilde{c}(z_1,k,z_2)$. On the other hand $\gamma$ is a smooth function, which can be numerically Fourier-transformed to $\tilde{\gamma}(z_1,k,z_2)$. Working in the Fourier-space with $\tilde{c}$ and $\tilde{\gamma}$ on the transformed inhomogeneous OZ equation \eqref{FT OZ equation planar eq2} yields updated functions $\tilde{c}^{\,'}$ and $\tilde{\gamma}^{\,'}$. The only quantity of interest is the latter. This can then be (mixed and) back-transformed to the real-space, where the closure relation will give back an updated $c^{\,'}$. This scheme is to be repeated until convergence.

\subsubsection{Implementation of the force-DFT}

For the considered two-dimensional planar geometry, the YBG equation \eqref{YBG equation}, which can first be re-expressed as
\begin{align*}
0 &= \nabla_{\vec{r}_1} \ln\!{\left( \rho(\vec{r}_1)\right)} + \nabla_{\vec{r}_1} \beta V_{\text{ext}}(\vec{r}_1)  \\
&\quad+ \frac{1}{ \rho(\vec{r}_1)} \!\int\! d \vec{r}_2 \,  \rho^{(2)}(\vec{r}_1,\vec{r}_2) \nabla_{\vec{r}_1} \beta \phi(r_{12}),
\end{align*}
yields
\begin{align}\label{YBG equation planar}
0 &= \frac{d\ln\!{\left(\rho(z_1)\right)}}{d z_1}  + \frac{d\beta V_{\text{ext}}(z_1 )}{d z_1} \\
&\quad + \frac{2}{\rho(z_1)} \int_{0}^{+\infty} \!\! dx_2 \int_{-\infty}^{+\infty} \!\! dz_2 \,  \rho^{(2)}(z_1,x_2,z_2 ) \frac{d\phi(r_{12})}{d z_1}, \notag
\end{align}
where we recall that $r_{12}\!=\!\sqrt{x_2^2 + (z_1-z_2)^2}$ and,
for $r_{12}\geq d=1$, we have that
\begin{equation}
\frac{d \phi(r_{12})}{d z_1} = \frac{(z_1-z_2)}{r_{12}} \frac{d \phi(r_{12})}{d r_{12}},
\end{equation}
with
\begin{equation}
\frac{d \phi(r_{12})}{d r_{12}} = - \phi(r_{12}) \left( \alpha + \frac{1}{r_{12}} \right).
\end{equation}
Integration of equation \eqref{YBG equation planar} from $0$ to $z$ and exponentiating then leads to
\begin{align}\label{force-DFT EL equation}
\rho(z) &= \rho_{0} \, e^{-\beta V_{\text{ext}}(z)} \rho_{\text{exp}}(z)
\end{align}
where we defined, for conveniency,
\begin{align}
\hspace*{-0.3cm}\rho_{\text{exp}}(z) \equiv e^{\int_0^z dz_1 \frac{2}{\rho(z_1)}
\int_{0}^{+\infty} \!\! dx_2 \int_{-\infty}^{+\infty} \!\! dz_2 \,  \rho^{(2)}(z_1,x_2,z_2 ) \frac{d\phi(r_{12})}{d z_1}}\!\!\!\!\!
\end{align}
and, more importantly, an integration constant
\begin{equation}\label{force-DFT normalization}
\rho_{0} \!\equiv\! \frac{\langle N \rangle}{\int_{-\infty}^{\infty} \! dz \, e^{-\beta V_{\text{ext}}(z)} \rho_{\text{exp}}(z)}, 
\end{equation}
for a given average number of particles per unit length, $\langle N \rangle\!=\!\int_{-\infty}^{\infty} dz \,\rho(z)$.

Equation \eqref{force-DFT EL equation} is the `EL-like' equation to be solved numerically (using Picard iteration) to obtain the force-DFT density profile for two-dimensional soft repulsive particles in planar geometry.

\section{Results}

\begin{figure*}
\includegraphics[width=0.9\linewidth]{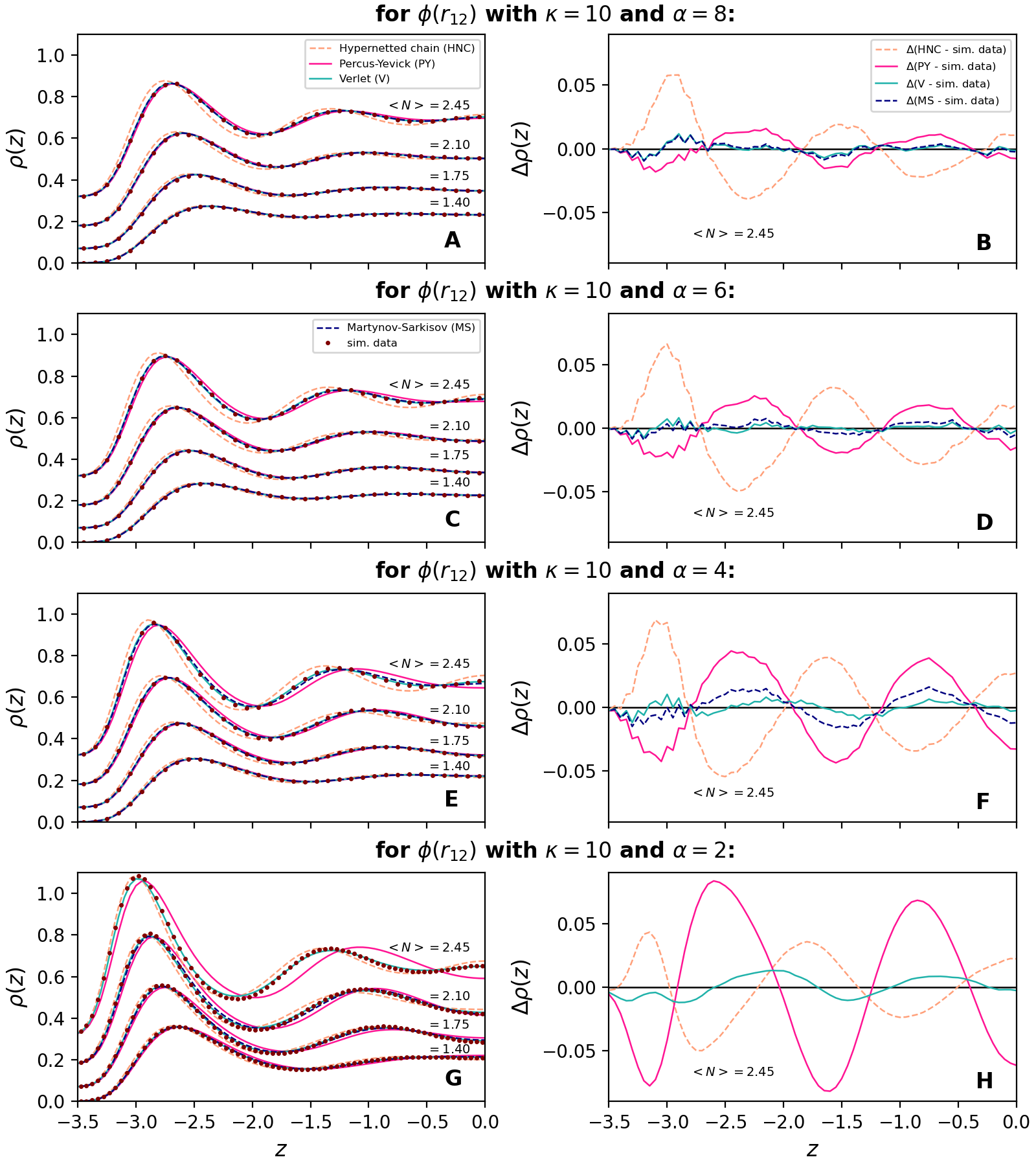}
\caption{ \textbf{Force-DFT with different closures for particles with different softness.}
We show (half of) the density profiles for a system of soft particles between two soft walls, as given by equations \eqref{HCY potential} and \eqref{external potential}, respectively. Each line corresponds to a different softness of the particles, while the external potential remains fixed. The first column shows the (shifted) density profiles, $\rho(z)$, and the second column the difference between the obtained density profiles and the simulation data, $\Delta \rho(z)$.
The different closures used are given as
{\red dashed light orange lines for the Hypernetted-chain \eqref{HNC closure},} 
solid pink lines for Percus-Yevick \eqref{PY closure}, solid seagreen lines for Verlet \eqref{V closure} and dashed navy blue lines for Martynov-Sarkisov \eqref{MS closure}. The density profiles generated by Brownian dynamics simulations is given as maroon dots.
} 
\label{fig equilibrium}
\end{figure*}

{\red
In the following  we will compare the outcomes of the (first principles) force-DFT promoted in this work with Brownian dynamics (BD) simulation data. For more details on the simulation, we refer the reader to Appendix~\ref{appendix details on BD sim data}.
}

In Fig.~\ref{fig equilibrium} we present numerical results for our chosen two-dimensional system of 
{\red softened disks} \eqref{HCY potential} between soft walls \eqref{external potential}.
The first column corresponds to the density profiles. Since they are mirror-symmetric about $z\!=\!0$, we only show half of them.
The profiles, evaluated at different average number of particles per unit length $\langle N \rangle\!=\!1.40,1.75,2.10$ and $2.45$, are also shifted vertically as follows:
{\red
\begin{equation}
\rho(z)\!+\rho_{\text{shift}},
\end{equation}
with $\rho_{\text{shift}}\!=\!0.00, 0.07, 0.18$ and $0.32$, respectively,} 
to enhance clarity by avoiding curve overlaps.
In the second column, we show the difference between the density profiles obtained by force-DFT and the simulation data, $\Delta \rho(z)$.
The force-DFT results are displayed in different colors depending on the used closure, namely
{\red dashed light orange lines for the Hypernetted-chain \eqref{HNC closure},}
solid pink lines for Percus-Yevick \eqref{PY closure}, solid seagreen lines for Verlet \eqref{V closure} and dashed navy blue lines for Martynov-Sarkisov \eqref{MS closure}. The data generated by BD
simulations are given as maroon dots.
Each of the four sets of panels in Fig.~\ref{fig equilibrium} corresponds to a different particle softness, i.e. $\alpha\!=\!8,6,4$ or $2$.
The external potential \eqref{external potential}, however, remains fixed.

The density profiles get more densly packed as we lower the $\alpha$-value used in the interparticle pair potential 
{\red (recall that lower values of $\alpha$ generate a longer 
range Yukawa repulsion)}. 
On the other hand, for a given $\phi$, an increase of the average number of particles per unit length also yields a higher packing.
Therefore, the most oscillatory density profile shown in Fig.~\ref{fig equilibrium} is the curve presented
{\red
in panel G
}
at $\langle N \rangle\!=\!2.45$ and the least oscillatory one is the curve in panel A at $\langle N \rangle\!=\!1.40$.
{\red
Since all closures used in this work correctly 
reproduce the exact low-density limit, there is no surprise that the force-DFT density curves coincide for the lower density profiles shown and only become significantly different as the packing grows.
}

{\red

For each of the panels in Fig. \ref{fig equilibrium} we increase $\langle N\rangle$ for a given value of the softness parameter, 
$\alpha$, and observe an increase in the packing oscillations of the one-body density. 
In order to better appreciate the packing level of the considered systems we have performed BD simulations of the corresponding bulk systems to roughly estimate the density, $\rho_b^{\text{fr}}$, at which a freezing transition is to be expected.  
As $\alpha$ is decreased the Yukawa tail in $\phi$ becomes longer ranged and thus the particles effectively take more space. 
This suggests that $\rho_b^{\text{fr}}$ should decrease as $\alpha$ is reduced. 
From our \textit{bulk} simulations we estimate the freezing densities to be at 
$\rho_b^{\text{fr}}\!=\!0.80, 0.775, 0.75, 0.70$, corresponding to area fractions $0.63, 0.61, 0.59, 0.55$, for $\alpha\!=\!8, 6, 4, 2$ and $\langle N\rangle\!=\!2.45$.
However, since in Fig. \ref{fig equilibrium} we are dealing with a highly inhomogeneous system, strongly confined between soft walls and with no flat `plateau' density-value even at the center of the slit, it is hard to draw any straightforward conclusions from simulations 
of a bulk system.
Nevertheless, our bulk freezing estimates can at least be used 
to give some feeling about whether the parameters we employ constitute a demanding test of the various force-DFT closures.  

From the simulation density profiles we read-off the values at 
the center of the slit, $\rho(0)$, and at the dominant 
peak close to the wall, $\text{max}\left[\rho(z)\right]$. 
We find $\rho(0)\!=\!0.385, 0.372, 0.357, 0.333$ 
and $\text{max}\left[\rho(z)\right]\!=\!0.542, 0.57, 0.638, 0.766$
for $\alpha\!=\!8, 6, 4, 2$. 
These values enable us to estimate that the most densely packed 
curves (those for $\langle N\rangle\!=\!2.45$) 
in Fig.~\ref{fig equilibrium} have values of $\rho(0)$ which are 
$48\%, 48\%, 47.6\%, 47.6\%$ of the freezing density 
and values of $\text{max}\left[\rho(z)\right]$ which are
$67.75\%, 73.55\%, 85.07\%, 109.43\%$ of the freezing density, for  
 $\alpha\!=\!8, 6, 4, 2$.
Although the values of $\rho(0)$ are all well below bulk freezing, 
the peak values lie very close to freezing for the smaller values of $\alpha$. 
It is important to remember that the 
self-consistent solution of force-DFT requires solution of the 
\textit{inhomogeneous} OZ equation to determine the 
inhomogeneous two-body correlations at all locations in the 
slit. In the vicinity of the main density peak, close to the wall, 
this clearly involves solving the OZ equation at a very high 
local density, comparable to, or even exceeding, in magnitude the bulk freezing 
density. 
We would thus claim that, within this spatial region, the self-consistent solution of force-DFT places considerable stress on 
the chosen closure relation and really tests the quality of 
the approximation. 
Since the force-DFT is a nonlocal theory, a good result for 
the entire density profile can only be obtained if the 
two-body closure is capable of handling the regions in which the 
density is highest. 
This is the `bottleneck' of the whole calculation.  
We thus conclude that the density profiles presented for 
$\langle N\rangle\!=\!2.45$ consitute demanding state 
points for testing our theory and represent 
quite `highly packed' situations, despite the fact that the density 
at the center of the slit is not particularly large.
Given these general considerations, 
we will now proceed to discuss the profiles in detail.

As we can already see in panel A, for the largest $\alpha$-value considered, $\alpha\!=\!8$, the density profiles with $\langle N \rangle\!=\! 2.10$ and $2.45$ generated using the PY and HNC closures differ slightly from those generated by the V or MS closures.  
The differences between the predictions of each approximation and the simulation data can be seen more clearly 
in panel B.
While the V and MS closures remain very similar and agree well with the simulation data, the PY profile shows some deviations even at these moderate packings. 
The HNC approximation performs worst, as could be anticipated for this short range Yukawa repulsion.

Although the outcomes of the V and MS closures are almost identical in the case $\alpha\!=\!8$, they start to differ ever so slightly from each other for $\alpha\!=\!6$, as shown in panels C and D. However, they both remain in excellent agreement with the simulation data.
In contrast, the PY force-DFT profile now exhibits more noticable deviations from the simulation data. 
We observe that for $\langle N \rangle\!=\! 2.45$ the PY density profile has the wrong shape: at $z\!=\!0$ the curve decreases to a local minima while the V, MS and simulation profiles all show a local maxima at this location. 
For this value of $\alpha$ the HNC closure becomes quite 
unreliable, with somewhat phase-shifted oscillations 
visible for the higher packings. 

For the value $\alpha\!=\!4$ the Yukawa repulsion begins 
to slowly change the character of the interparticle potential from hard-disk-like to a more softened interaction.  
The trends apparent in the profiles are generally similar 
to those for $\alpha\!=\!6$, but now become more pronounced 
as the particles effectively take more space and are thus 
more strongly subject to packing constraints. 
The force-DFT profiles calculated using the V and 
MS closures are in very close agreement with the simulation data, although inspection of Panel E indeed reveals some 
small differences between the two. 
The PY profiles become quite inaccurate for 
$\langle N \rangle\!=\! 2.45$ and the `shape problem' (i.e.~the prediction of an unphysical minima at $z\!=\!0$) pointed out previously is exacerbated. 
The HNC gives a generally poor account of the simulation 
data, although for these strongly softened disks the predictions are not substantially worse than those of the PY theory. 
It thus appears that both of the `classic' liquid-state closures, PY and HNC, seem to have 
difficulties. 
Overall, from inspection of Panel F we can conclude that 
the V closure starts to emerge as the most reliable of the considered approximations. 

In Panels G and H we consider profiles calculated for strongly softened disks with a Yukawa 
parameter $\alpha\!=\!2$. 
This is the most demanding packing situation we will consider. 
At $\langle N \rangle\!=\! 2.45$ the PY closure is in serious error and erroneously predicts strongly 
phase-shifted oscillations relative to the simulation data. 
In contrast, the HNC approximation, although still exhibiting notable errors, gives a fairly reasonable 
account of the general shape of the simulation density profile. 
The improving performance of the HNC closure, and the worsening performance of the PY closure, in moving 
from $\alpha\!=\!8$ to $\alpha\!=\!2$ is consistent with experience from bulk integral equation studies, 
for which the PY and HNC are known to perform best for strongly repulsive and softly repulsive interaction potentials, respectively. 
However, we note that it is not to be taken for granted that this past experience gained from the investigation 
of bulk pair correlations will neccessarily carry over to the present situation, for which a chosen closure 
forms part of a self-consistent iteration scheme to determine the one-body density. 
Profiles from the MS closure are absent from Panels G and H at $\langle N \rangle\!=\! 2.45$, since the theory becomes unstable and we 
could not obtain converged solutions. This difficulty is related to an intrincic limitation of the MS 
theory, namely the presence of a square root in the approximation to the bridge function, equation 
\eqref{MS bridge}, as we mentioned previously. 

To conclude, the data presented in Fig. \ref{fig equilibrium} demonstrates that 
the density profiles generated by the V closure are the most accurate, sitting right under the simulation data points, and stably so for the full range of parameters explored.
The MS force-DFT profiles are just behind, at least for the cases where a converged solution could be obtained, with density profiles very close to those generated by the V closure. Only at high packing do the two closures show some small but visible discrepencies, with the V closure proving more accurate in all cases.

The profiles generated using the PY and HNC closures are clearly inferior to the V and MS results. 
As the level of softness increases the PY closure presents a decrease in quantitative accuracy and, 
more seriously, the shape of the curves has the tendency to become wrong at higher packing, missing the 
maxima at $z\!=\!0$.
The HNC approximation is generally the worst of the closures considered, but redeems itself as the softness 
becomes large (see Panels G and H), where it performs far better than the PY. 

One should note that our harsh assessment of the PY and HNC closures is only a consequence of the excellently 
accurate predictions we obtain using the MS and V approximations. 
When viewed in the broader context of standard DFT predictions, where profiles are often only qualitatively 
predicted (see e.g.~\cite{articleDFTbetter}), even the PY and HNC results presented here are quite 
acceptable. 
Our recommendation for employing force-DFT to treat softened repulsive particles is to use the V closure (which is both accurate and stable) 
over all the others tested in this work. The numerical scheme to implement any of these closures is the exact 
same up to one single line of code. There is thus no reason not to choose the best.

}

{\red
\section{Structural inconsistency}

Thermodynamic inconsistency is a difficulty which arises in 
\textit{bulk} integral equation studies. 
If the bulk pair correlation functions are known, then the 
thermodynamic properties of the system can be calculated 
using one of three routes: the internal energy, the virial 
or the compressibility \cite{hansen06}. 
Within an exact approach all three yield perfectly consistent results. 
However, when the pair correlation functions are only 
known approximately, as is usually the case when 
using closures of the OZ equation, then the three routes 
are no longer equivalent \cite{caccamo}. 
Although this feature certainly complicates the 
application of integral equation approximations, it has 
been exploited as a means to optimise closure approximations
by incorporating adjustable parameters which can be fixed 
to enforce partial consistency (see Refs. 
\cite{scozaWilding,scozaOther} for a particularly successful example of this approach).  

In the present work we employ integral equation closures 
of the \textit{inhomogeneous} OZ equation and couple this to the 
YBG equation \eqref{YBG equation} to obtain a closed theory.
The YBG equation provides 
an exact relation between $\rho^{(2)}$ and $\rho$. 
One can argue that using this equation is the most intuitive choice, since the YBG force-balance emerges 
in a very natural way when addressing problems of colloidal dynamics \cite{tschopp4,tschopp5,tschopp6}. 
The YBG is, however, not the only exact equation connecting the one-body density to the two-body correlation 
functions.   
For example, the Lovett-Mou-Buff-Wertheim (LMBW) equation \cite{LMBW1,LMBW2}, given in general by
\begin{align} \label{LMBW equation}
0 = \nabla_{\vec{r}_1}  \rho(\vec{r}_1) &+  \rho(\vec{r}_1) \nabla_{\vec{r}_1} \beta V_{\text{ext}}(\vec{r}_1) \\
&- \rho(\vec{r}_1) \!\int\! d \vec{r}_2 \,  c(\vec{r}_1,\vec{r}_2) \nabla_{\vec{r}_2} \rho(\vec{r}_2), \notag
\end{align}
and yielding
\begin{align}\label{LMBW equation planar}
\!\! \frac{d\rho(z_1)}{d z_1}  &=  
2 \rho(z_1) \int_{0}^{+\infty} \!\! dx_2 \int_{-\infty}^{+\infty} \!\! dz_2 \, 
c(z_1,x_2,z_2 ) \frac{d\rho(z_2)}{d z_2} \notag \\
&\quad - \rho(z_1) \frac{d\beta V_{\text{ext}}(z_1 )}{d z_1}
\end{align}
in planar geometry, makes an explicit link 
between $c$ and $\rho$.
As mentioned above, both the YBG equation \eqref{YBG equation} and the LMBW equation \eqref{LMBW equation}
are \textit{formally} exact. 
However, these equations cannot be simultaneously satisfied when approximations are involved.  
Our strategy to reveal the presence of such a `structural inconsistency' is to take the converged $\rho$ and $c$ from the force-DFT (in which the YBG one-body equation is used in the self-consistancy loop) and input these to the right-hand side of equation \eqref{LMBW equation planar} to generate a new output $\rho_{\text{LMBW}}'$. This will then be compared with $\rho_{\text{YBG}}'\!\equiv\!\rho'$, the numerical derivative of the density profile obtained from our self-consistent force-DFT calculation. Results are shown in Fig. \ref{fig structural inconsistency}.
The numerical derivative of the converged density profile (as given in Fig. \ref{fig equilibrium}), $\rho_{\text{YBG}}'$, is shown as a solid line colored differently according to the used closure, namely light orange for the Hypernetted-chain \eqref{HNC closure}, pink for Percus-Yevick \eqref{PY closure}, navy blue for Martynov-Sarkisov \eqref{MS closure} and seagreen for Verlet \eqref{V closure}.
For each case, the corresponding $\rho_{\text{LMBW}}'$ is given by the dashed olive green curve.
As a reference, the numerical derivative of the simulation data density profile (also given in Fig. \ref{fig equilibrium}) is shown as maroon dots.
Each column in Fig. \ref{fig structural inconsistency} tests a different particle softness, by varying the $\alpha$-value in the interaction potential \eqref{HCY potential}.
There is one row of Panels per closure relation.

\begin{figure*}
\includegraphics[width=0.8\linewidth]{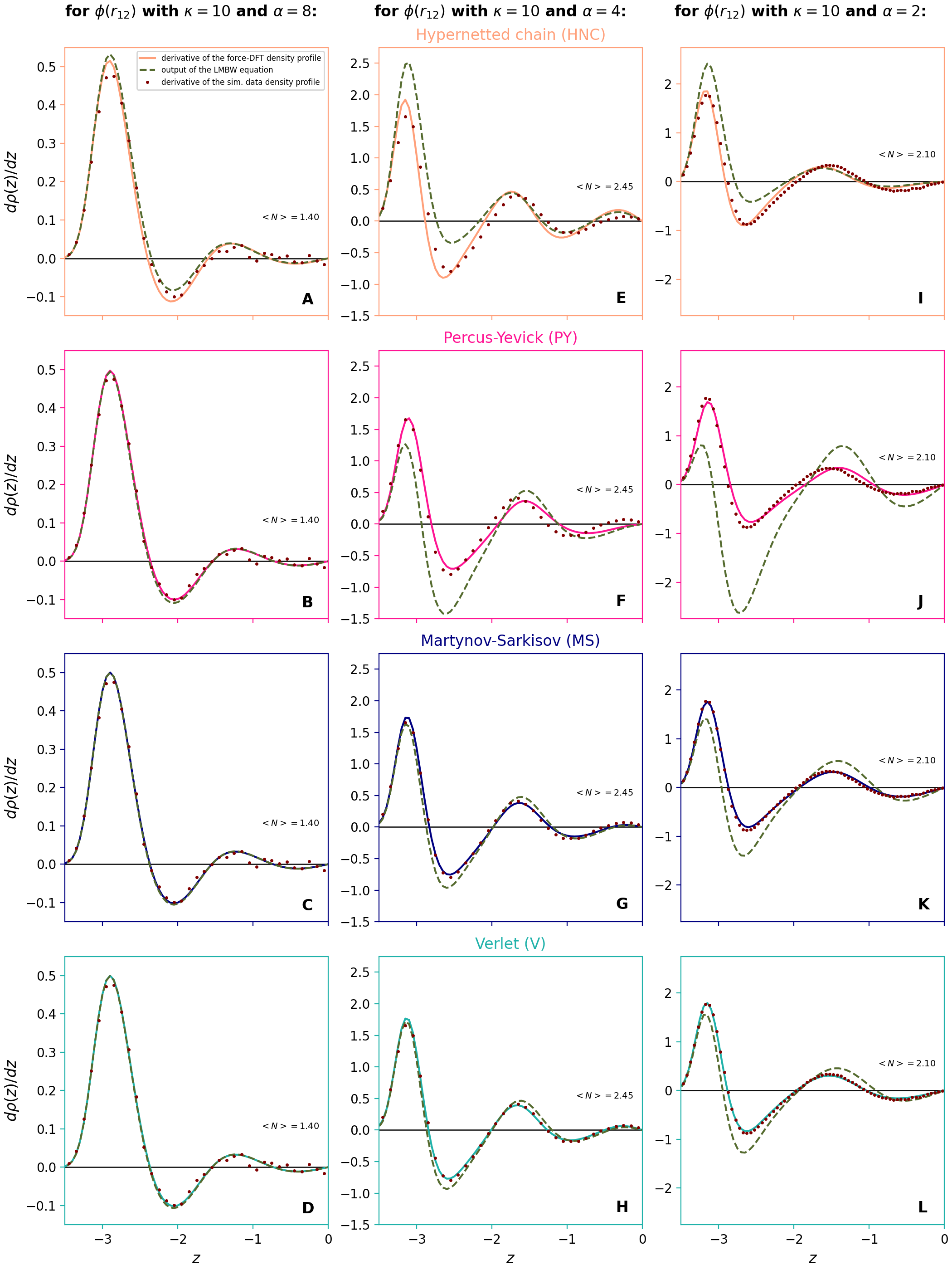}
\caption{ \textbf{Structural inconsistency.}
{\red
For each panel, taking the corresponding density profile obtained via force-DFT (Fig. \ref{fig equilibrium}) and putting it and its, also shown, numerical derivative (solid colored line - light orange, pink, navy blue and seagreen, depending on the applied closure) into the right-hand side of the planar LMBW equation \eqref{LMBW equation planar} yields the dashed 
olive green curve. The discrepency between those two latter curves is a test of the structural inconsistency. The numerical derivative of the simulation data density profile (Fig. \ref{fig equilibrium}) is also given as maroon dots, as a reference.
Each column represents another type of particle softness, by varying the $\alpha$-value in the interaction potential \eqref{HCY potential}, while each row gives results for the same closure relation, namely
the Hypernetted-chain \eqref{HNC closure},
Percus-Yevick \eqref{PY closure}, Martynov-Sarkisov \eqref{MS closure} and Verlet \eqref{V closure}.
}
} 
\label{fig structural inconsistency}
\end{figure*}

Let us begin by considering the least demanding situation, namely $\alpha\!=\!8$ and $\langle N \rangle\!=\! 1.40$, and following 
down the first column of Panels to compare the predictions of the various closure relations. 
Unsurprisingly, the HNC approximation already shows notable structural inconsistency. 
The PY closure improves upon this, but is still inferior to the results from the MS and V closures, which are almost 
perfectly self-consistent.
The second column shows results for $\alpha\!=\!4$ and $\langle N \rangle\!=\! 2.45$. 
Here we can see clearly the failings of both the HNC and PY closures, due to the high density but intermediate 
particle softness. 
It is interesting to note that the numerical derivative of the self-consistent force-DFT density profile, $\rho_{\text{YBG}}'$, agrees quite well in both cases with the simulation data, whereas the output $\rho_{\text{LMBW}}'$ deviates significantly from them, albeit in opposite directions. The result for $\rho_{\text{LMBW}}'$ in Panel E overestimates the force-DFT prediction, while the one in Panel F underestimates it. 
Finally, results for $\alpha\!=\!2$ and $\langle N \rangle\!=\! 2.10$ (which is the highest packing for which the MS scheme still has a solution) are shown in the last column of Fig. \ref{fig structural inconsistency}. At this higher particle softness the HNC closure gives better results than the PY, for which we observe in Panel J a very poor level of structural inconsistency with $\rho_{\text{LMBW}}'$ greatly undershooting both the force-DFT and BD predictions. 
The HNC result for $\rho_{\text{LMBW}}'$ shown in Panel I still overshoots, but is considerably better than in the case 
$\alpha\!=\!4$ and $\langle N \rangle\!=\! 2.45$ (Panel E). 
The level of structural inconsistency from the HNC is now comparable to that of the MS closure, given in Panel K. 
The best closure relation, generating the least amount of structural inconsistency is clearly the V approximation.

We observe that all results for $\rho_{\text{LMBW}}'$ underestimate the force-DFT and BD simulation results, with the exception of those obtained using the HNC approximation. 
This may reflect the differing types of system for which these closures were intended. 
The PY, MS and V were targeted at treating hard particles, while the HNC was focused rather on soft penetrable ones. 
The trends of the inconsistency appear to be linked to this fundamental difference.

In this work we have chosen to focus on purely repulsive interactions for which the only phase transition is freezing and 
we have chosen parameters so as to stay below this value. 
For such systems we have shown that the best closures (MS and V) exhibit a quite acceptable level of 
structural inconsistency, which provides a reassuring quality check. 
However, it is well-known that thermodynamic (and, thus, surely structural) inconsistency will play a more important role for 
systems exhibiting a liquid-vapour phase transition. 
In such cases much care in both implementation and interpretation is required. 
For most closures there exist unphysical `no-solutions' regions of the thermodynamic parameter space which are closely 
associated with the existence of spinodal lines and which often prevent integral equation theories being applied for the 
determination of the phase diagram 
\cite{BraderThermophysics}.    
For these reasons we defer to future work the delicate analysis of the force-DFT and structural inconsistency for attractive systems.

}

\section{Discussion}

In this paper we have applied force-DFT to calculate the one-body density in (two-dimensional) planar geometry. 
By generalizing known bulk closures to solve the inhomogeneous OZ equation and coupling this with the 
exact YBG equation we have shown that this alternative route to force-DFT 
can provide a robust and accurate method for predicting the 
one-body density profile from first principles. 
For the model interaction potential considered here the absence 
of a good quality approximation to the Helmholtz free energy functional would not permit results of such high precision to 
be obtained using the traditional implementation of DFT. 

The versatility of force-DFT lies in its ability to exploit 
the large body of existing knowledge on closures of 
the bulk OZ equation, many of which have been assessed and 
well documented in a recent article by 
Pihlajamaa and Janssen \cite{Janssen}. 
In contrast, reports in the literature concerning use of generalized closures of the 
inhomogeneous OZ equation are quite rare  
and limited almost entirely to studies 
of three-dimensional hard-spheres using 
the PY approximation 
\cite{AttardSpherical,nygard1,nygard2,tschopp1}, although there are exceptions (see for example References \cite{Brader2008Struc-885,kovalenko}). 
The investigation of generalized closures within 
force-DFT thus provides great opportunities 
for future studies of inhomogeneous fluids and offers the 
potential to accurately calculate density profiles from 
first principles in cases where standard (free energy-based) 
DFT approximations can only make predictions of qualitative accuracy. 
In short, inhomogeneous OZ closures are powerful, but often overlooked, non-perturbative 
approximations which can be usefully applied to many problems 
of interest.  

One may well ask why such integral equation closures, which were developed to treat bulk systems, are equally applicable to 
closing the inhomogeneous OZ equation, and thus the force-DFT. 
The explanation can be found in the origin of these approximations as diagrammatic cluster expansions \cite{StellCluster,hansen06,AttardBook}. 
Many of the familiar integral equation closures, such as the 
Percus-Yevick, Hyper-Netted-Chain etc., can be derived as partial resummations of Mayer cluster expansions to yield closed form expressions \cite{mcquarrie}. 
The form of the diagrammatic expansions remains the same 
in the inhomogeneous case as in bulk; factors of $\rho_b$ 
associated with integration field points in bulk systems 
are simply replaced by factors of $\rho(\vec{r})$ in the inhomogeneous case. 
This correspondence means that any resummation scheme 
employed in bulk will be of equal validity in the presence 
of spatial inhomogeneity. 
Closures developed to treat the bulk might therefore be expected to work just 
as well for closure of the inhomogeneous OZ equation. 
The results shown in Figure \ref{fig equilibrium} 
certainly support this expectation.

Of the {\red four} approximations considered, the Verlet 
closure was found to produce the best results when compared 
with BD simulation data. 
In bulk, Verlet found that when applied to three-dimensional hard-spheres the closure \eqref{V closure} produced radial 
distribution functions which yielded a high level of 
consistency for the equation of state when comparing 
the pressure obtained via the virial and compressibility 
routes (roughly on the $1\%$ level over the entire range 
of bulk density up to crystallization) \cite{verlet_closure}. 
Given these findings it is perhaps not surprising, although 
certainly gratifying, to find that application of the 
inhomogeneous Verlet approximation to close the 
force-DFT yields density profiles in excellent agreement 
with simulation data. 
{\red However, it is important to} point out that it is not obvious  
that residual errors in the two-body Verlet closure would 
not become amplified as a consequence of coupling to the one-body density via the YBG relation. 
That this is not the case speaks for the stability 
of the force-DFT scheme. 

In his paper on integral equation closures in which the bulk 
version of equation \eqref{V closure} was introduced, Verlet also 
proposed a slightly modified approximation, incorporating an 
adjustable parameter. 
For given values of the bulk density and temperature the value of 
this parameter is 
adjusted to enforce perfect virial-compressibility consistency 
on the pressure. 
This raises the question of whether it could be possible to 
generalize the Verlet-modified closure to the inhomogeneous 
case by replacing the adjustable parameter with a spatially dependent function, which could 
then be uniquely 
determined by generalized consistency requirements on the 
inhomogeneous one- and two-body density. 
Similar considerations could be applied to inhomogeneous generalizations of other `thermodynamically consistent' bulk closures, such as the well-known Rogers-Young approximation 
\cite{rogersyoung}. 
{\red In Fig. \ref{fig structural inconsistency} we examined 
the structural inconsistency of the various closures 
considered in this work. Although the original version of 
the Verlet closure proved impressively consistent, there 
is still clearly room for optimisation. } 

In the present work we have focused on (two-dimensional) planar geometry, 
however the use of generalized integral equation 
closures within force-DFT would also be of great interest 
when applied to (three-dimensional) spherical or (two-dimensional) polar geometries. 
This would allow for `test particle' calculations in 
which the external potential is simply a fluid particle 
fixed at the coordinate origin. 
The \textit{bulk} radial distribution function $g(r)$ can 
then be obtained from the inhomogeneous (radially symmetric) 
one-body density by using the Percus identity, 
$g(r)\!=\!\rho(r)/\rho_b$. 
With this method any given closure of the bulk OZ equation 
can be automatically upgraded in accuracy:  
the inhomogeneous generalization of the closure is used 
within the force-DFT scheme to obtain the radial density 
profile and, thus, a new and improved estimate for $g(r)$.    
For example, we anticipate that applying the Verlet closure 
\eqref{V closure} to force-DFT for three-dimensional hard-spheres would 
produce bulk radial distribution functions (and thus pressures) {\red from the test particle method}
which are more accurate than those reported in \cite{verlet_closure}, where the closure was directly applied 
to the \textit{bulk} OZ equation.

Finally, despite the many appealing features of force-DFT, 
the absence of a free energy functional is a disadvantage when 
attempting to study phase transitions.  
The spectre of thermodynamic 
inconsistency, familiar from bulk liquid-state 
integral equations, would undoubtedly rise once more 
{\red via structural inconsistency} when applying generalized closures to the inhomogeneous OZ equation. 
Finding ways to deal with such issues within force-DFT 
remains a topic for future study.

\appendix

\section{Derivation of the Fourier-transformed OZ equation}\label{appendix section FT OZ}

We start by rewriting the general form of the inhomogeneous OZ equation \eqref{OZ equation} for our specific geometry, namely the two-dimensional planar case as illustrated in figure \ref{fig sketch geometry}.
This yields
\begin{align}\label{OZ equation planar}
& h(z_1,x_2,z_2) = c(z_1,x_2,z_2) \\
&\quad+ \!\int_{-\infty}^{+\infty}\!\! dx_3 \int_{-\infty}^{+\infty}\!\! dz_3\, h(z_1,x_3,z_3)\,  \rho(z_3) \, c(z_3,x_{32},z_2),\notag
\end{align}
where we recall that $x_{1i}\!=\!x_{i}$.
Using the definition of the back-Fourier transformation, the integral term can then be rewritten as
\begin{multline*}
\frac{1}{(2\pi)^2}\int_{-\infty}^{+\infty}\!\! dx_3 \int_{-\infty}^{+\infty}\!\! dz_3\, \rho(z_3) \int_{-\infty}^{+\infty}\!\! dk \, \tilde{h}(z_1,k,z_3) \, e^{ikx_3} \\
\times \int_{-\infty}^{+\infty}\!\! dk' \, \tilde{c}(z_3,k',z_2)\, e^{ik'x_{32}}.
\end{multline*}
Since $x_{32}\!=\!x_2-x_3$ and $\int_{-\infty}^{+\infty}\!\! dx_3 \, e^{i(k-k')x_3}\!=\!2\pi \delta(k-k')$, it reduces to
\begin{equation*}
\frac{1}{2\pi} \int_{-\infty}^{+\infty}\!\! dz_3\, \rho(z_3) \int_{-\infty}^{+\infty}\!\! dk' \, \tilde{h}(z_1,k',z_3)\, \tilde{c}(z_3,k',z_2)\, e^{ik'x_2}.
\end{equation*}
Plugging this last expression into the OZ equation \eqref{OZ equation planar} and then applying a Fourier transform on the $x_2$-coordinate yields
\begin{align*}
&\tilde{h}(z_1,k,z_2) = \tilde{c}(z_1,k,z_2) +  \int_{-\infty}^{+\infty}\!\! dx_2
\Bigg(
 \frac{1}{2\pi} \int_{-\infty}^{+\infty}\!\! dz_3\, \rho(z_3)
 \Bigg.
 \\& \quad \quad \Bigg. \times \int_{-\infty}^{+\infty}\!\! dk' \, \tilde{h}(z_1,k',z_3)\, \tilde{c}(z_3,k',z_2)\, e^{ik'x_2}
\Bigg) 
  \, e^{-ikx_2},\notag
\end{align*}
Using once more that $\int_{-\infty}^{+\infty}\!\! dx_2 \, e^{i(k'-k)x_2}\!=\!2\pi \delta(k'-k)$, we recover equation \eqref{FT OZ equation planar eq1} in the main text.

{\red

\section{Brownian dynamics simulations}\label{appendix details on BD sim data}

The simulations for the density profiles of disks confined between walls were performed in the canonical ensemble using the LAMMPS package~\cite{LAMMPS2022} and a 
two-dimensional simulation box of size 
$L_{x} \!\times\! L_{z} \!=\! 100 \!\times\! 11 d^2$, with periodic boundaries along the $x$-axis. 
The pairwise interactions between particles are modeled via a pseudo-hard-disk potential, which provides a continuous approximation to the idealized hard-disk interaction \cite{jover2012}, plus 
an additional repulsive Yukawa potential, as given by 
equation \eqref{HCY potential}. 
Each system is equilibrated for $10^7$ integration time steps, followed by a production run of $2\!\times\!10^8$ steps. The particle positions are recorded every $5\!\times\!10^3$ steps.

Note that, the channel width has been changed from $8 d$ (as used in force-DFT) to $11 d$, since in simulations, a hard wall cannot be defined at a specific point. Instead, we set harmonic walls with thickness of $1.5$ such that particles at $z\!=\!-4$ and $4$ see a potential similar to a hard wall. Particles are not able to pass positions $z\!=\!-3.5$ and $3.5$ and the density is always zero at $z\!=\!-4$ and $4$.

To determine the bulk freezing density, we performed BD simulations for hard disks in a two-dimensional simulation box of the same size as used previously, but with periodic boundary conditions applied along both axes. Other computational details remain consistent with those used for disks confined between walls, except that no walls are present and the external potential, $V_{\text{ext}}$, is absent.
As the bulk density of disks increases, the first peak of the radial distribution function, $g(r_{12})$, shifts towards smaller disk-disk separation distances, $r_{12}$, while its height decreases. As the bulk density is increased beyond the freezing density the peak position no longer moves, but instead its height increases.
We identify this threshold as the freezing density of the bulk system, $\rho_b^{\text{fr}}$.

}

\bibliographystyle{unsrtnat}
\bibliography{bibliography_tschopp}

\end{document}